\documentclass{moriond}
\def\Journal#1#2#3#4{{#1} {\bf #2}, #3 (#4)}

\usepackage{lineno}
\usepackage{amsmath}
\newcommand\tablesze{\fontsize{8pt}{8pt}\selectfont}

\begin{document}
\vspace*{4cm}
\title{Search for Dilepton Resonances with the ATLAS Detector and Run 2 Data}
\author{Aaron White\\
        On behalf of the ATLAS Collaboration}
\address{University of Michigan, Department of Physics}

\maketitle\abstracts{
A search for resonances in the dielectron and dimuon mass spectra from 250 GeV to 6 TeV is presented. The data were recorded during Run 2 of the LHC by the ATLAS experiment using proton-proton ($pp$) collisions with a center-of-mass energy of $\sqrt{s} = 13$ TeV. The integrated luminosity of the data corresponds to 139 fb$^{-1}$. The background models are a functional form fit to the data, and generic signal shapes are used to represent various models with different resonant widths and masses. No significant deviation from the expected background is observed and 95\% confidence level upper limits are set on the fiducial cross-section times branching ratio for models of various widths. For benchmark models, limits are converted to lower limits on the resonance mass and reach 4.5 TeV for the E$_6$ motivated $Z'_\psi$ boson.
}

\section{Introduction}
The dielectron and dimuon final states have been useful for a number of discoveries including the $J/\psi$ and $\Upsilon$ mesons, as well as the $Z$ boson. These discoveries have aided the establishment of the Standard Model (SM), and they motivate this search for resonances with a pole mass between 250 GeV and 6 TeV. This proceeding summarizes the results which are described in detail in the paper \emph{Search for high-mass dilepton resonances using 139 fb$^{-1}$ of $pp$ collision data collected at $\sqrt{13}$ TeV with the ATLAS detector} \cite{full} by the ATLAS Collaboration \cite{atlascolab}.

Various models predict spin-1 vector bosons which decay to dileptons. The Sequential Standard Model predicts a $Z'_{SSM}$ boson associated with a new U(1) gauge group, with the same fermion couplings as the SM $Z$ boson \cite{zmssm}. The E$_6$-motivated Grand Unification model predicts another $Z'_\chi$ and $Z'_\psi$ bosons also associated with an additional U(1) symmetry \cite{ze6}. The Heavy Vector Triplet model predicts a $Z'_{HVT}$ boson which is a neutral member of a new SU(2) gauge group \cite{hvt}. This search is also sensitive to spin-0 resonances such as that predicted by the Minimal Supersymmetric SM \cite{mssm}, or spin-2 resonances such as those predicted in the Randall-Sundrum model \cite{rs}.

This study presents a search for a new resonance decaying to either two electrons or two muons in 139 fb$^{-1}$ of data collected in center of mass energy $\sqrt{s} = 13$ TeV $pp$ collisions at the LHC. Previous searches have been conducted by the ATLAS and CMS collaborations \cite{zpatlas,zpcms} showing no excess beyond the background, and leading to lower limits of up to 3.8 TeV for the mass of the $Z'_\psi$ boson. This study benefits from a larger data sample by a factor of four, the use of a functional form to model the backgrounds instead of a simulated background, and the use of generic signal shapes that can readily be interpreted for different models.

\section{Detector and Data}
ATLAS is a multipurpose detector built in a cylinder around the LHC beam axis. Starting from the center, the Inner Detector (ID) help reconstruct particle tracks bent by a 2 T axial magnetic field for $|\eta|<2.5$. Outside of this, electromagnetic and hadronic calorimeters measure energy deposits covering $|\eta|<4.9$. The outermost layer consists of a Muon Spectrometer (MS) covering $|\eta|<2.7$ and incorporating a toroidal magnetic field.

The dataset used for this study was collected between 2015 and 2018 during the LHC Run 2, and includes only events during stable beam conditions and nominal detector operation. For the dielectron channel, events were recorded using the ``very loose'' or ``loose'' identification criteria, and a transverse energy threshold ($E_\text{T}$) threshold depending on the year. For the dimuon channel, events were recorded using one of two single muon triggers: any muon with $p_\text{T}>50$ GeV, or isolated muon with $p_\text{T}>26$ GeV \cite{elecmuon}. The integrated luminosity has been determined to be $139.0\pm2.4$ fb$^{-1}$ \cite{lumi}.

Electron candidates are reconstructed from ID tracks that match with energy deposits in the electromagnetic calorimeter. The ``medium'' working point is used which has an efficiency above 92\% for $E_\text{T}>80$ GeV. Muon candidates are also reconstructed using ID and matching tracks from the MS. The ``high-$p_\text{T}$'' working point is used which requires muon hits in each of the three MS stations and vetoes regions with suboptimal alignment, with an efficiency of 69\% at 1 TeV. There are additional cuts on longitudinal ($z_0$) and transverse ($z_0$) impact parameters, $\eta$, and $E_\text{T}(p_\text{T})$ for electrons (muons). The selection requirements for muon and electron candidates is shown in Table \ref{objs}. Additional details on cuts to reduce background are described in the full paper.

Events are required to contain at least two same-flavor leptons. Dimuons are required to be oppositely charged, but same-sign dielectrons are permitted. In events with multiple dilepton candidates, electrons with the highest $E_\text{T}$ and muons with the highest $p_\text{T}$ are used. In events with both dielectron and dimuon pairs, the dielectron pair is preferred due to the higher efficiency.

\begin{table}
\caption[]{Electron and muon selection requirements}
\label{objs}
\vspace{0.4cm}
\begin{center}
{\tablesze
\begin{tabular}{|l|c c|}\hline
Selection & Electrons & Muons\\\hline\hline
Working point & medium & high-$p_\text{T}$\\
Pseudo-rapidity & $|\eta|<1.37$ or $1.52<|\eta|<2.47$ & $|\eta|<2.5$\\
$E_\text{T}$ & $E_\text{T}>30$ GeV & \\
$p_\text{T}$ &              & $p_\text{T}>30$ GeV\\
Impact parameter & $|d_0/\sigma(d_0)|<5$ & $|d_0/\sigma(d_0)|<3$ \\
Impact parameter & \multicolumn{2}{c|}{$|z_0 \sin\theta|<0.5$ mm}\\
\hline\end{tabular}}
\end{center}
\end{table}

\section{Signal and Background Modeling}

\begin{figure}
\begin{minipage}{0.50\linewidth}
\centerline{\includegraphics[width=1\linewidth]{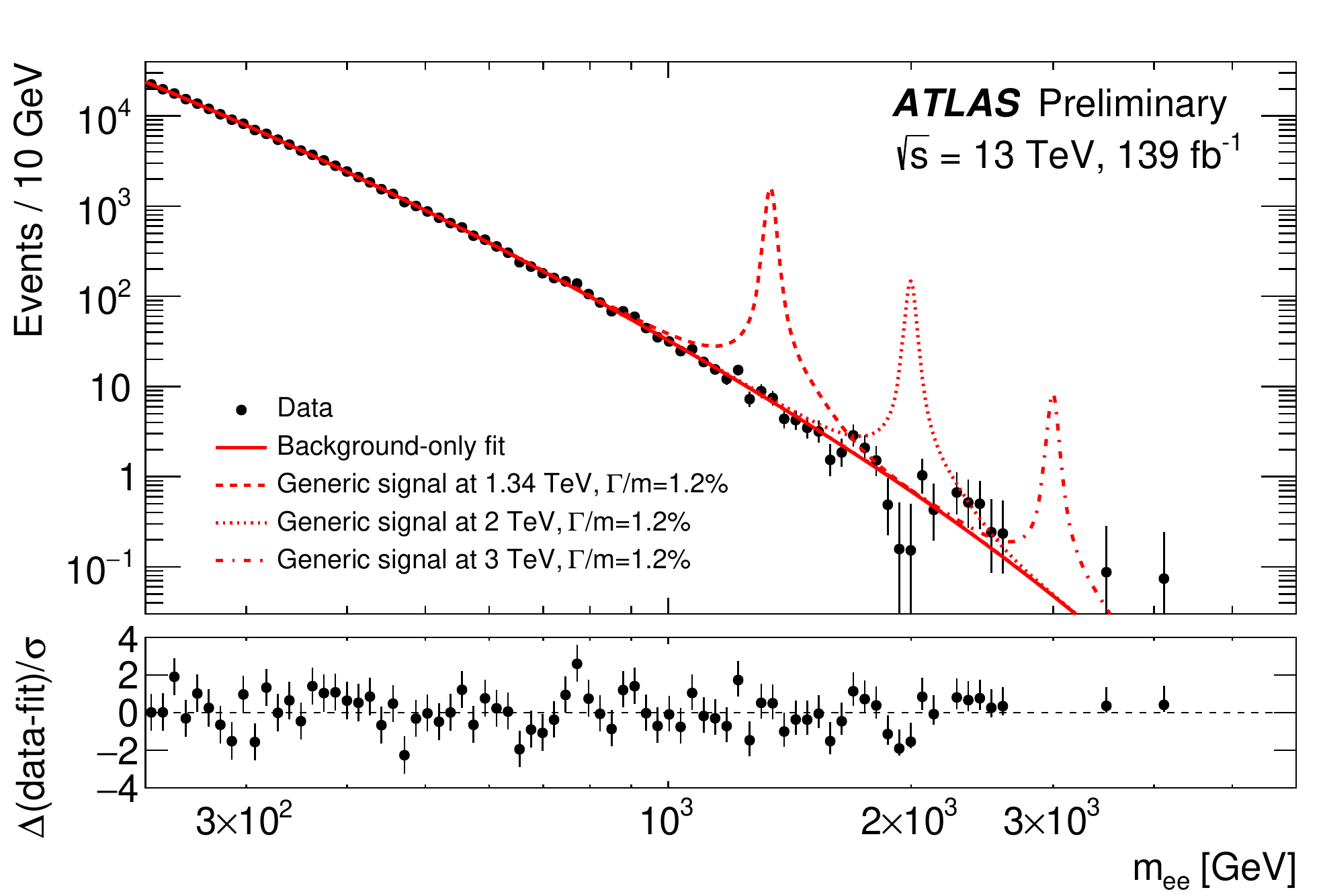}}
\end{minipage}
\hfill
\begin{minipage}{0.50\linewidth}
\centerline{\includegraphics[width=1\linewidth]{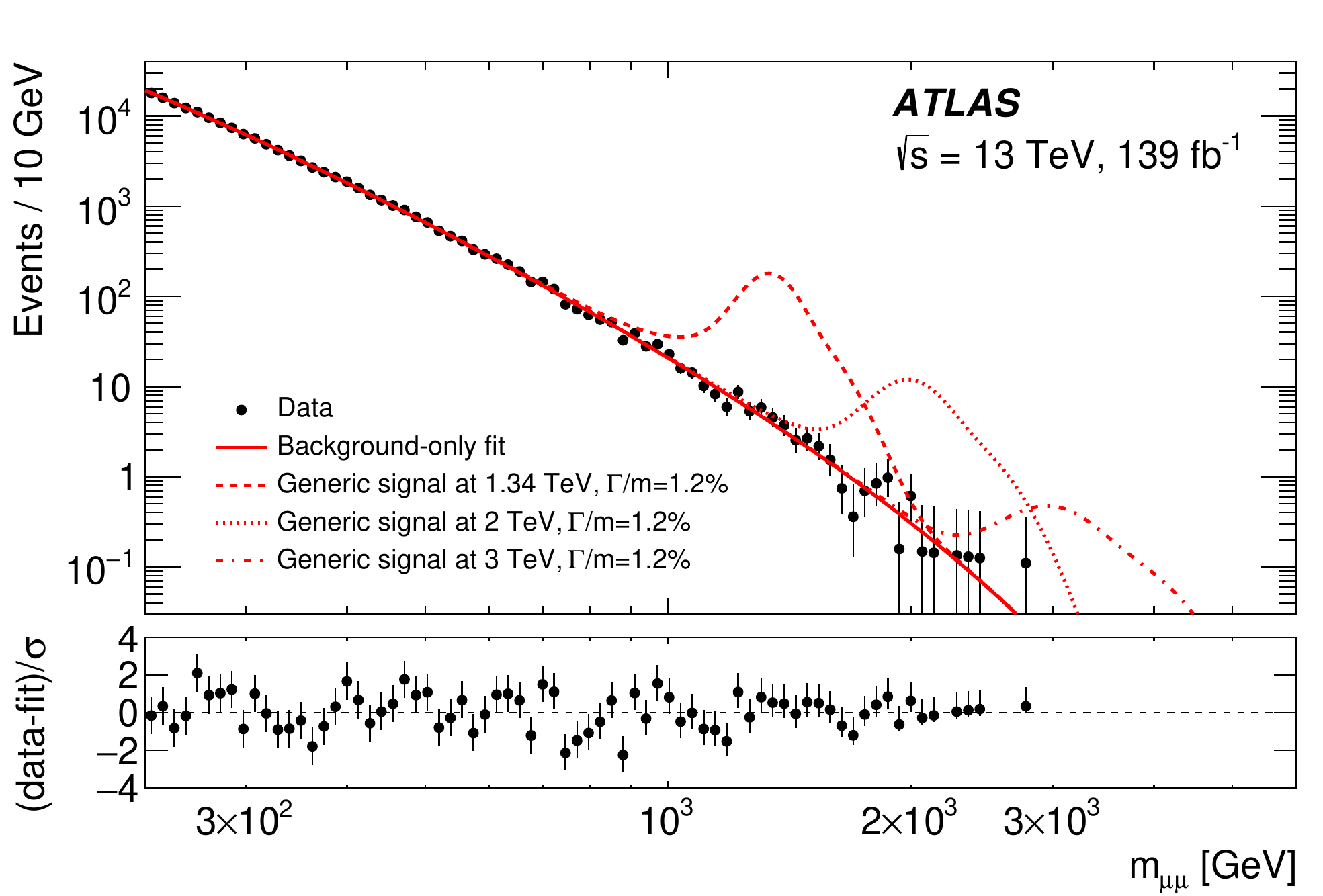}}
\end{minipage}
\caption[]{Dielectron invariant mass spectrum (left) and dimuon invariant mass spectrum (right). On top of the smoothly falling background, some generic signal shapes are drawn in dotted lines.}
\label{mass}
\end{figure}

This analysis searches for resonant shapes in the smoothly falling dilepton spectra as seen in Figure \ref{mass}. A smooth functional form plus a generic signal shape is fit to the data to model background and signal, respectively. The background is modeled using a smooth functional form. The choice of the form is based on MC performance studies that attempted to minimize the spurious signal arising from the background model. The selected form for the background shown in Equation \ref{eq:spa} 

\begin{equation}
f(m_{ll})= a \cdot f_{\text{BW},m_Z,\Gamma_Z}(m_{ll}) \cdot (1-x^c)^b \cdot x^{\sum^{3}_{i=0} p_i \log^i(x)},
\label{eq:spa}
\end{equation}

\noindent where $x=m_{ll}/\sqrt{s}$, and parameters $b$ and $p_i$ are determined by the fit to data. Parameter $c$ is chosen to be 1 for the dielectron channel and 1/3 for the dimuon channel. $f_{\text{BW},m_Z,\Gamma_Z}(m_{ll})$ is the non-relativistic Breit-Wigner function with $m_Z$ and $\Gamma_Z$ as the $Z$ mass and width, respectively. $a$ normalizes the background function to the total number of background events.

The generic signal shape is a non-relativistic Breit-Wigner function to model the physical width of the resonance, convolved with the sum of a Gaussian and a crystal ball shapes to model detector resolution. The detector resolution is determined via a comparison between reconstructed and truth mass using MC simulation. The generic signal shape is a determined by a reconstructed mass $m_X$ and width $\Gamma_X$. The systematic uncertainty depends both on $m_X$ and $\Gamma_X$. A fiducial region is defined in order to interpret this generic shape for different models predicting a dilepton resonance. For a given model, each lepton must pass $|\eta|<2.5$ and $E_\text{T}(p_\text{T})>30$ GeV, and the dilepton mass must satisfy $m^\text{true}_{ll}>m_X-2\Gamma_X$ where $m^\text{true}_{ll}$ is the simulated dilepton mass at Born level before reconstruction. This reduces the impact from off-shell effects not modeled by the generic signal shape.


\section{Results}

\begin{figure}
\begin{minipage}{0.50\linewidth}
\centerline{\includegraphics[width=1\linewidth]{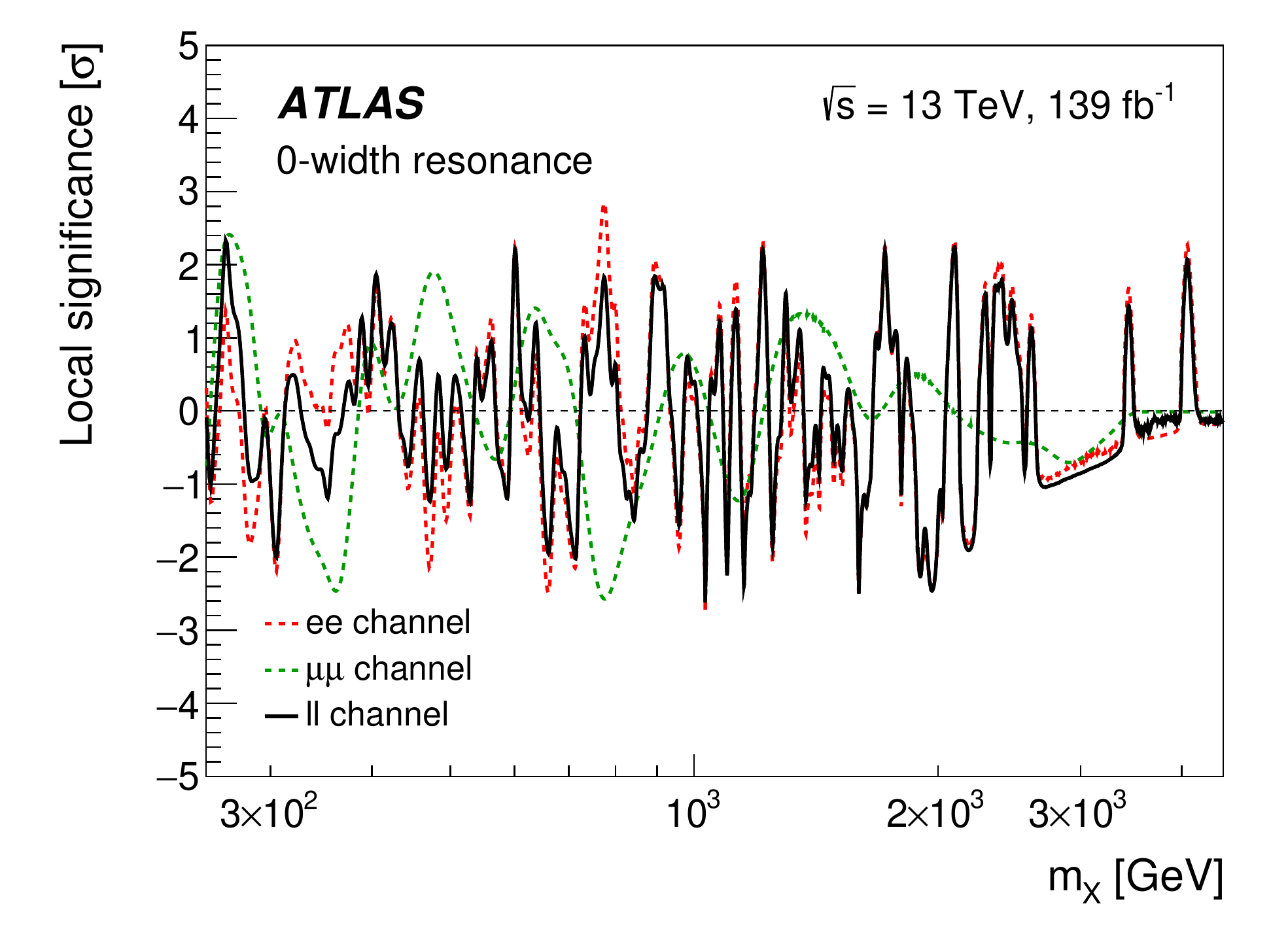}}
\end{minipage}
\hfill
\begin{minipage}{0.50\linewidth}
\centerline{\includegraphics[width=1\linewidth]{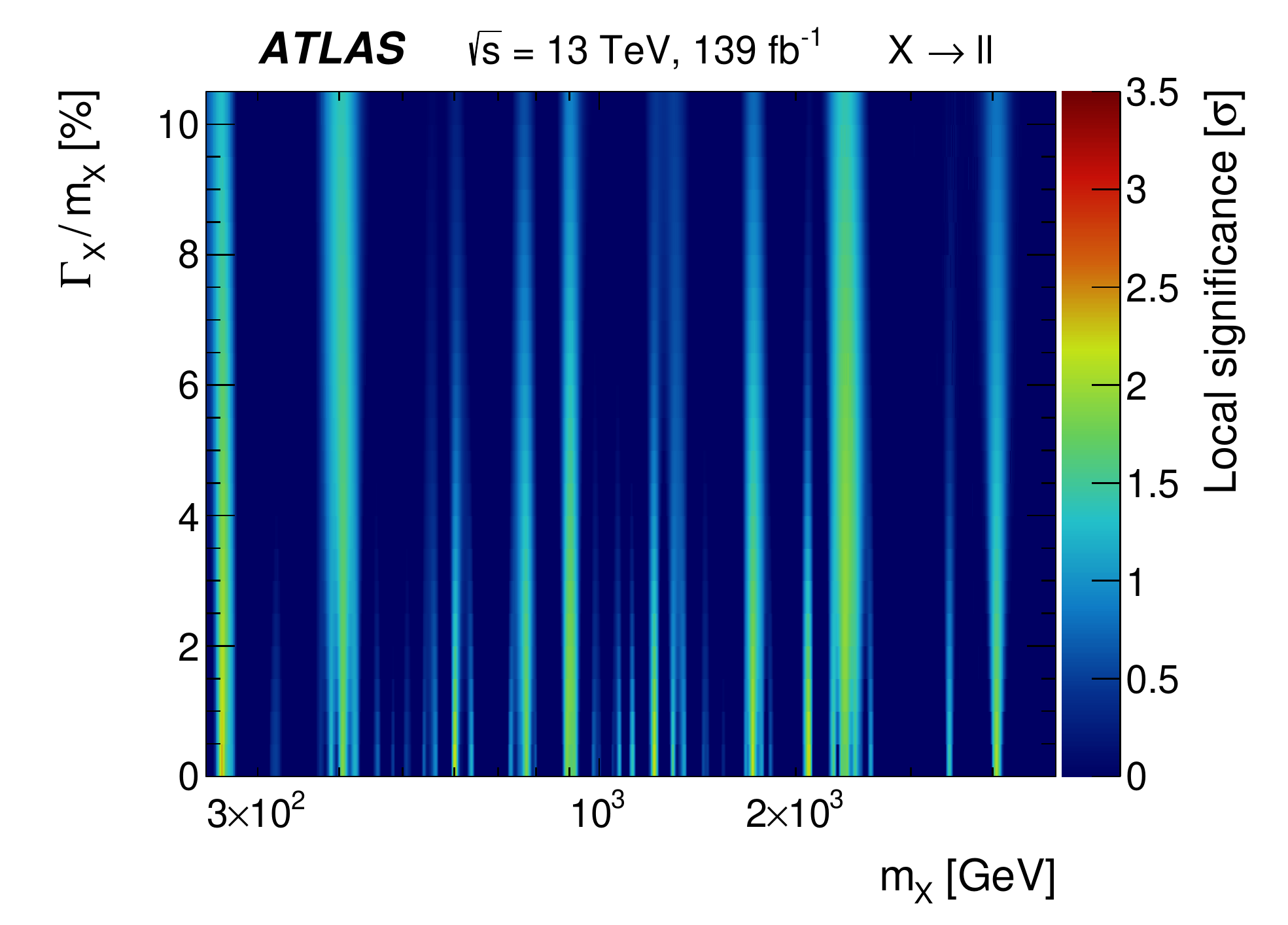}}
\end{minipage}
\caption[]{Probability that the spectrum is compatible with the background-only hypothesis for the dielectron, dimuon and combined dilepton channels. The local p0 is quantified in standard deviations $\sigma$. To the left, the zero-width significance scan in mass is shown. To the right, the vertical axis shows the scan repeated for various signal widths.}
\label{prob}
\end{figure}

The numbers of signal and background events are measured using a fit of the signal-plus-background model to the dilepton mass distribution. A background-only fit is also performed. Systematic uncertainties are taken into account via nuisance parameters constrained by either Gaussian (energy and momentum scale) or log-normal (efficiency and resolution) distributions in the likelihood. The spurious signal is represented by allowing a non-zero signal normalization under the background-only hypothesis. The dielectron and dimuon channels are fit separately, and then combined under a lepton-flavor universality assumption.

The invariant mass distributions are shown in Figure \ref{mass}. In the full Run 2 dataset, the event with the highest reconstructed mass is a dielectron candidate with $m_{ee}=4.06$ TeV, while the highest mass dimuon event has $m_{\mu\mu}=2.75$ TeV.

Signal models are considered for reconstructed masses from 250 GeV to 6 TeV in steps of 1 GeV, and for relative widths from 0\% to 10\% in steps of 0.5\%. The probability that the data are compatible with the background-only hypothesis as a function of pole mass for zero-width signals is shown on the left in Figure \ref{prob}. The 2D scan in pole mass and relative width is shown on the right. No significant excess is observed. For the zero-width case, the largest deviations from the background-only hypothesis for the dielectron, dimuon, and combined dilepton channels are observed at masses of 774 GeV, 267 GeV, and 264 GeV. These have global significances of 0.1$\sigma$, 0.3$\sigma$, and $\sim0$, respectively.

\begin{figure}
\begin{minipage}{0.50\linewidth}
\centerline{\includegraphics[width=1\linewidth]{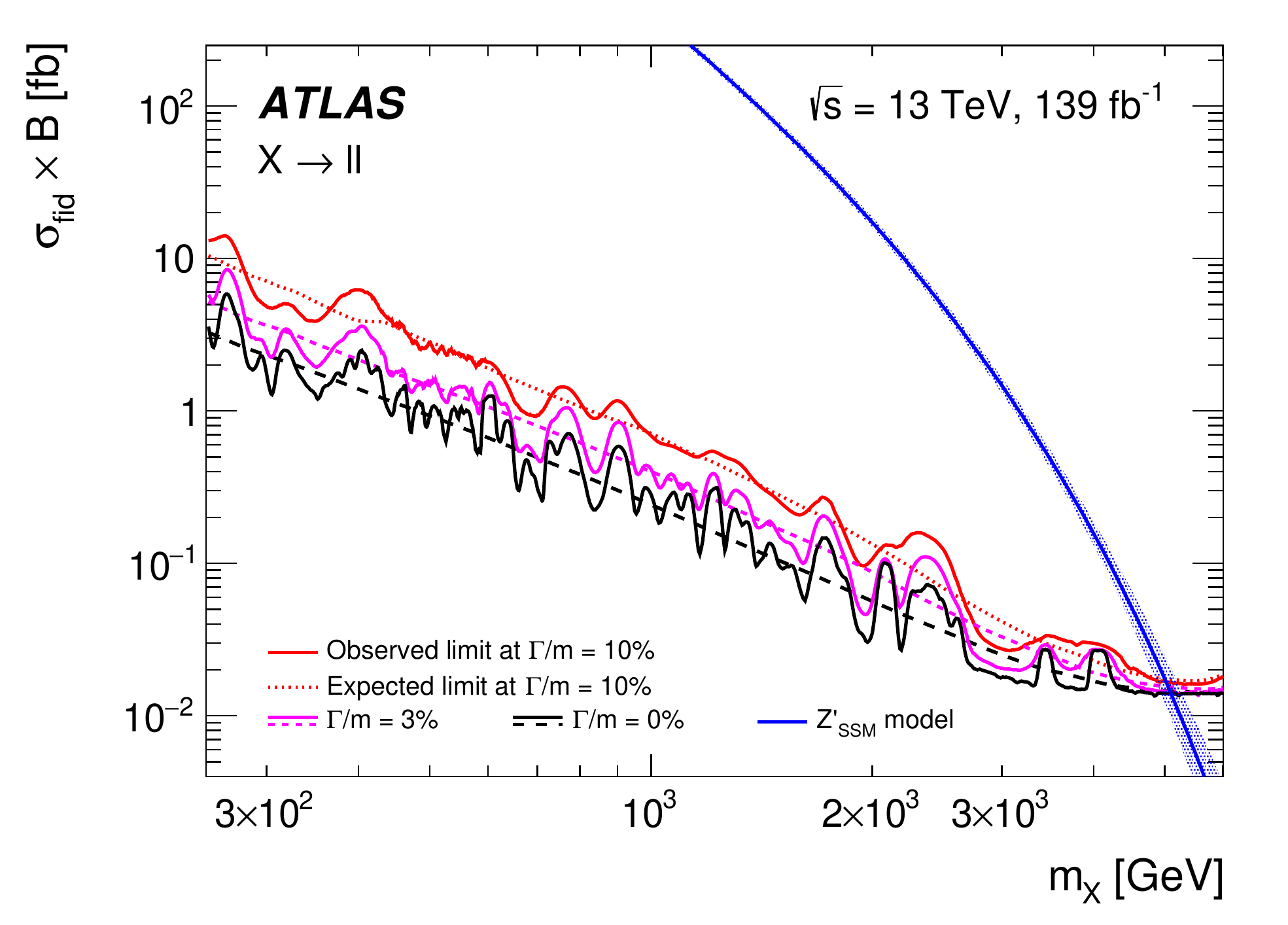}}
\end{minipage}
\hfill
\begin{minipage}{0.50\linewidth}
\centerline{\includegraphics[width=1\linewidth]{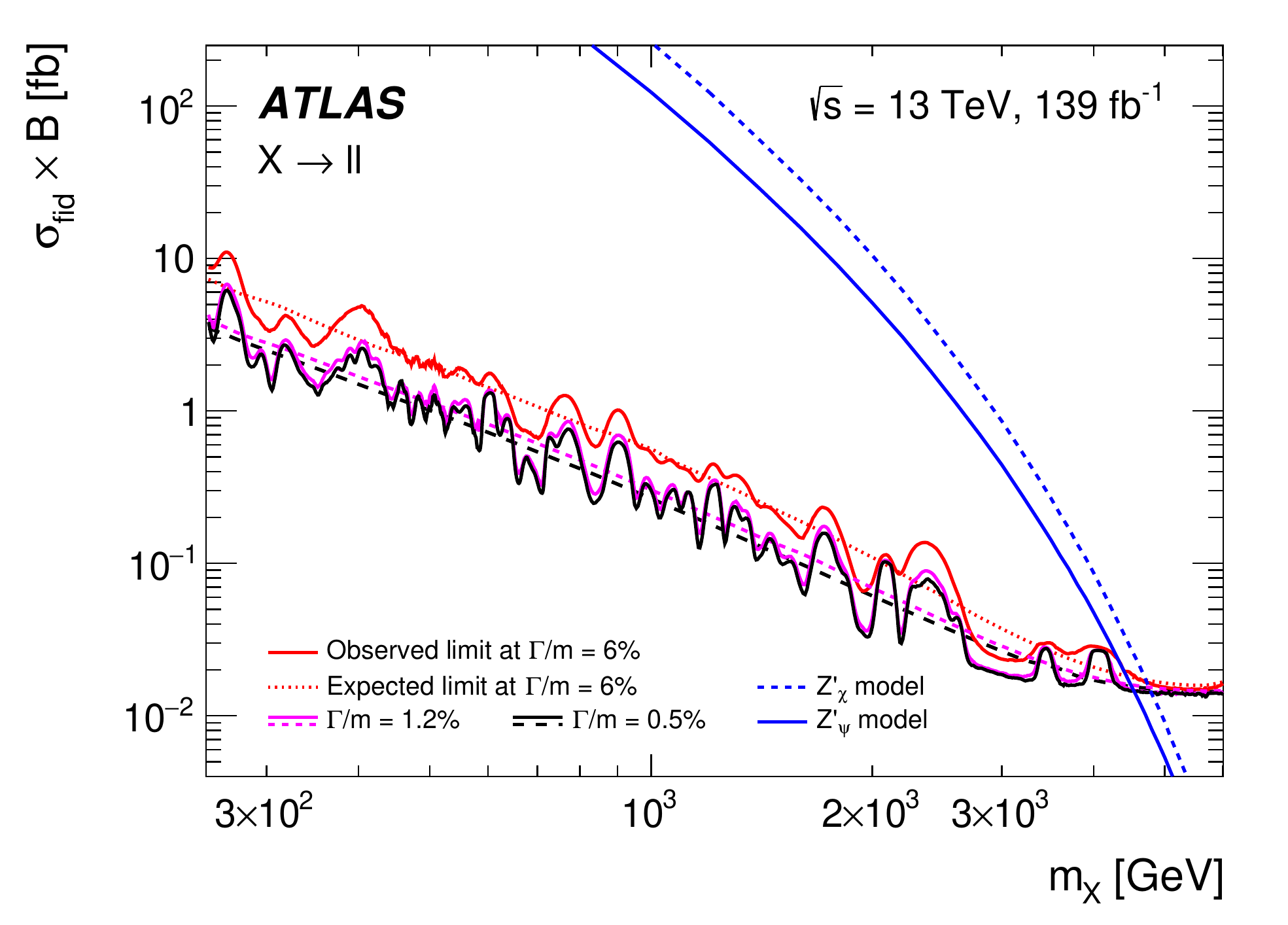}}
\end{minipage}
\caption[]{Upper limits at 95\% CL on the fiducial cross section times branching ratio to for the zero-width, 3\% and 10\% relative width signals as a function of pole mass. In blue are theoretical cross-sections for $Z'_\text{SSM}$, $Z'_\psi$, and $Z'_\chi$.}
\label{limits}
\end{figure}

The upper limits on the fiducial cross-section times branching ratio for the dilepton combination for several relative widths is shown in Figure \ref{limits}. Several model predictions are superimposed on top of the limits for $Z'_\text{SSM}$, $Z'_\psi$, and $Z'_\chi$ bosons. Masses below the intersection of the model prediction and the observed limit are considered excluded. The observed and expected lower limits are shown in Table \ref{lowerLims}. This result improves upon the previous exclusion by 500-800 GeV.

\begin{table}
\begin{center}
\begin{minipage}{0.50\linewidth}
\centerline{\includegraphics[width=1\linewidth]{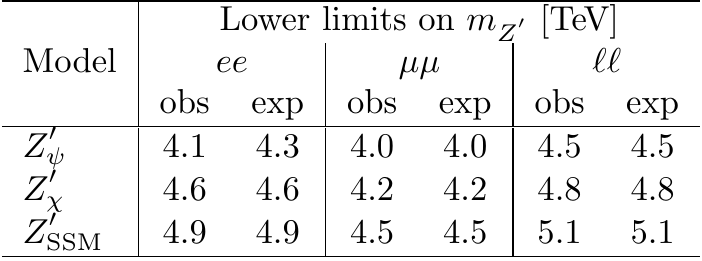}}
\end{minipage}
\end{center}
\caption[]{Observed and expected 95\% CL lower limits on $m_{Z'}$ for three models}
\label{lowerLims}
\end{table}

In summary, the ATLAS collaboration has searched for resonances in the dilepton mass spectra from 250 GeV to 6 TeV. The search covers the full $\sqrt{s}=13$ TeV Run 2 dataset corresponding to 139 fb$^{-1}$ collected at the LHC. The background estimate uses a functional form fit to data, and the search covers resonances with a variety of masses and relative widths. No significant excess is found above background, and limits are set on the fiducial cross-section times branching ratio for generic resonances with a relative natural width range from 0 to 10\%. Lower limits on several benchmark $Z'$ models are set, and limits on the heavy vector triplet model couplings are set and described in the full paper.

\break

\noindent Copyright 2019 CERN for the benefit of the ATLAS Collaboration. CC-BY-4.0 license.

\end{document}